\documentclass[letterpaper]{jpconf}
\usepackage{graphicx}

\begin{document}
\title{Getting Ready for Physics - Commissioning of the ALICE Experiment}

\author{H.R. Schmidt, for the ALICE Experiment
}

\address{ GSI Helmholtz Centre for Heavy Ion Research GmbH,  Darmstadt }

\begin{abstract}
ALICE at CERN-LHC is an experiment dedicated to the study of high-energy heavy-ion collision. In this paper we will briefly describe the experimental layout and  give an overview on the installation status of the ALICE detector components including an outlook towards the completion of the staged detector setup. We will review the commissioning of the detector components with emphasis on the central detectors, in particular the time projection chamber (TPC). The commissioning took place in summer 2008 awaiting the first beams from the LHC in the fall.  The result from commissioning are compared with the performance figures, which are outlined in the Physics Performance Reports \cite{PPR1, PPR2}.

\end{abstract}

\section{Introduction}
The physics of hot and dense matter has been investigated over the last two decades at the AGS \cite{AGS}, the SPS\cite{SPS} and at RHIC \cite{RHIC_PHOBOS, RHIC_BRAHMS, RHIC_STAR, RHIC_PHENIX}. The results obtained in recent years at RHIC have confirmed the previous SPS findings and, beyond that, given exciting new insights into the physics of strongly interacting matter. ALICE (A Large Ion Collider Experiment) is designed to study the physics of hot and dense matter and the transition to a quark-gluon plasma at the CERN Large Hadron Collider (LHC). The ALICE detector is optimized to cope with the highest particle multiplicities anticipated for PbÐPb collisions ($dN_{ch}/dy$ up to 5000). In addition to heavy systems, the ALICE Collaboration will study collisions of lower-mass ions, which are a means of varying the energy density, and protons (both pp and pA), which primarily provide reference data for the nucleus-nucleus collisions. The pp data will in addition allow for a number of genuine pp physics studies.

\section{Experimental Layout}
\begin{figure}
\centering
\includegraphics[width=12 cm,clip]{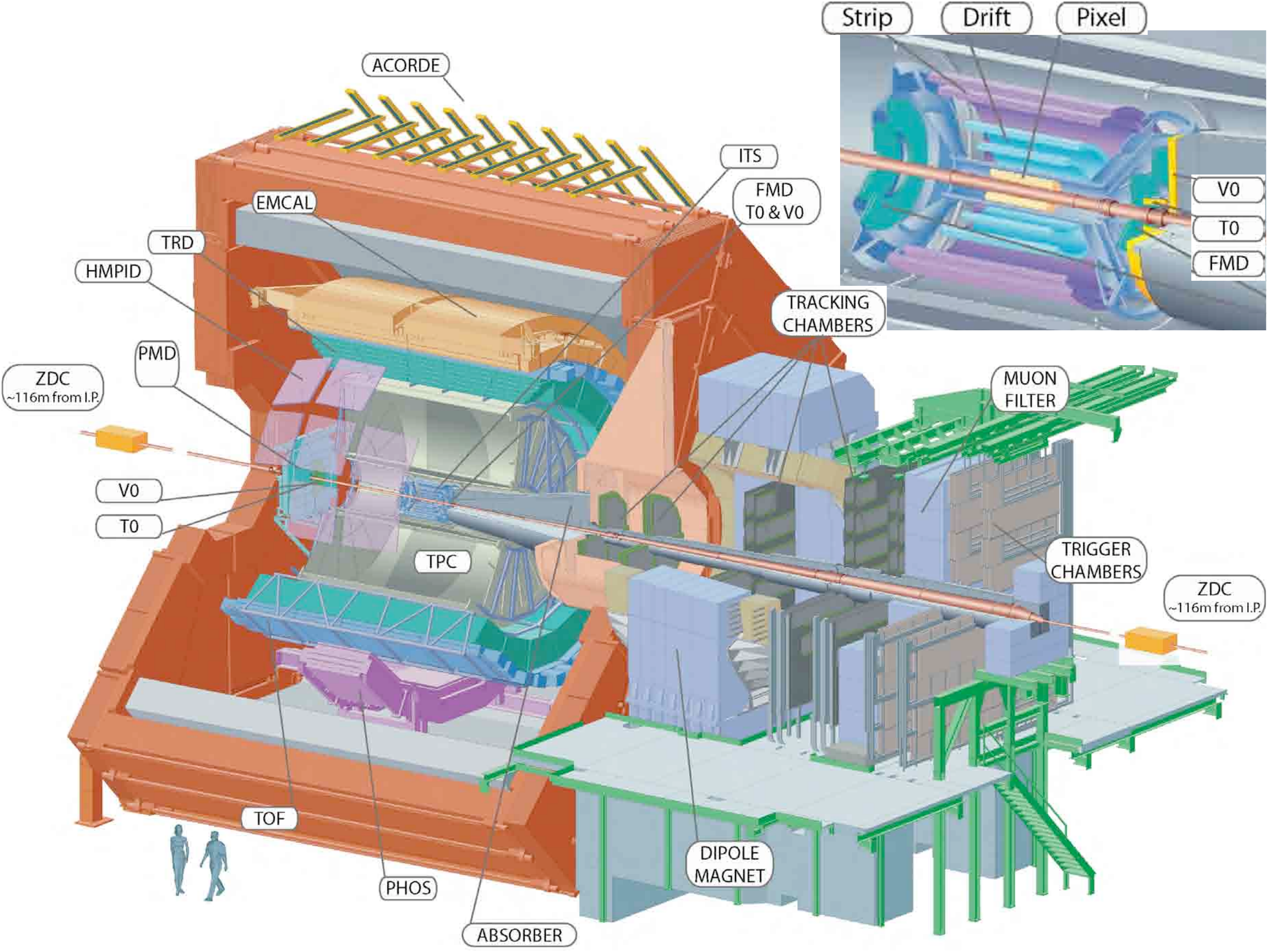}
\caption{The ALICE detectors.}
\label{ALICEDet}
\end{figure}

The detector (cf. Figure~\ref{ALICEDet}) consists of a central part, which measures event-by-event hadrons, electrons and photons, and of a forward spectrometer to measure muons. The central part, which covers polar angles from $45 {^\circ}$ to $135 {^\circ}$ ($|\eta| <0.9$) over the full azimuth, is embedded in the large L3 solenoidal magnet. The central barrel consists of: an Inner Tracking System (ITS) of high-resolution silicon detectors; a cylindrical Time-Projection Chamber (TPC),  a Transition Radiation Detector (TRD) and a Time-Of-Flight (TOF) detector.   A large lead-scintillator Electromagnetic Calorimeter covering  $\Delta \phi =110 {^\circ}$  and $|\eta| <0.9$ complements the central barrel of ALICE. 

The single-arm detectors PHOS , a lead-tungstate crystal electromagnetic calorimeter and HMPID,  a ring imaging Cherenkov hodoscope add to the central detectors. 

The forward muon arm (covering polar angles $\theta$ from  $171 {^\circ}$ to $178 {^\circ}$) consists of a complex arrangement of absorbers, a large dipole magnet, and 14 planes of tracking and triggering chambers. Several smaller detectors (PMD, ZDC, FMD, T0, V0) for global event characterization and triggering are located at forward angles. An array of scintillators (ACORDE) on top of the L3 magnet will be used to trigger on cosmic rays.

At the time of submitting this paper (May 2009)  the central detectors ITS, TPC and TOF are completely installed and are ready for beam in fall 2009. The TRD had 4 out of 18 super-modules installed and will be upgraded to 8 modules in summer 2009. Furthermore, the forward muon spectrometer as well as the central HMPID single-arm spectrometer are completely installed.  Both the scintillator-lead calorimeter  and the high resolution PHOS photon spectrometer are at present partially installed and are schedule to be completed in 2010/11. The installation of the Photon Multiplicity Detector PMD will be completed during 2009. The ZDC (Zero Degree Calorimeter) as well as the other trigger detectors are ready. A High Level Trigger System (HLT) will be fully operational in 2009.

Thus, at the foreseen restart of the  LHC in fall 2009 the ALICE detector will have full hadron and muon and partial electron and photon capability.

\begin{figure}
\centering
\includegraphics[width=14cm,clip]{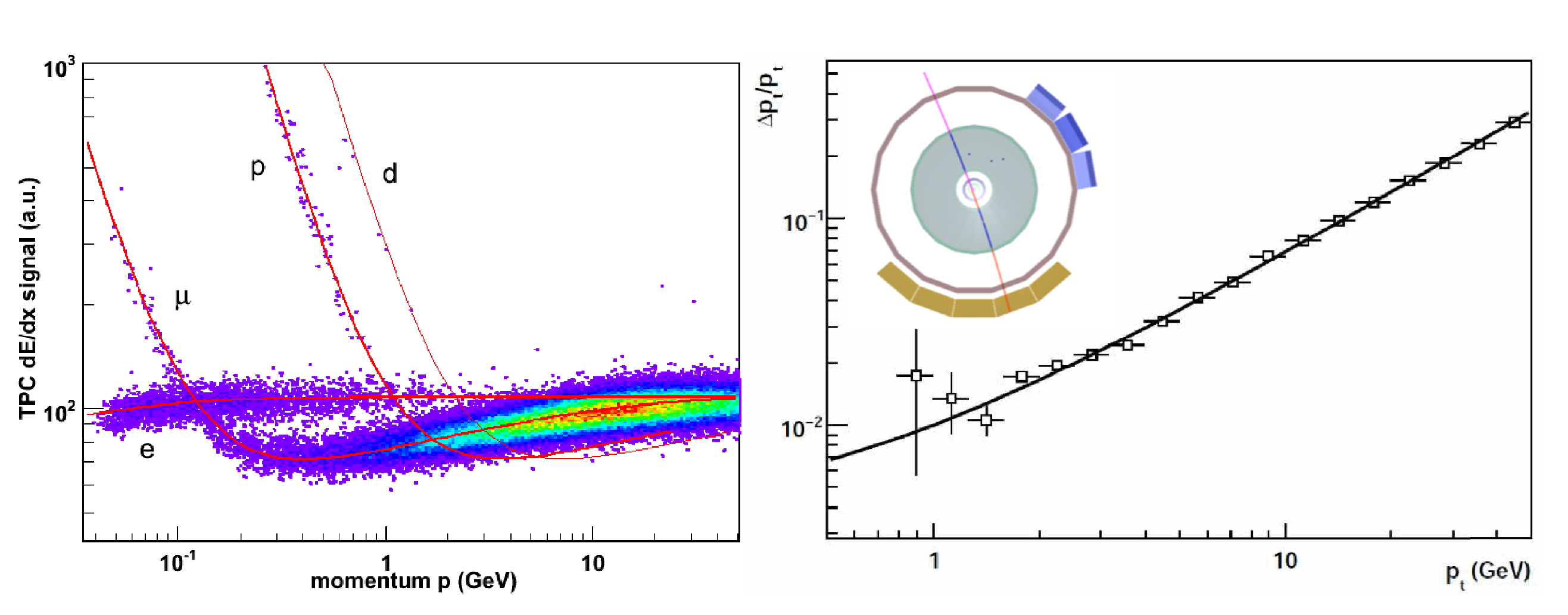}
\caption{Left: Differential energy loss dE/dx of charged tracks in the TPC as function of momentum together with Bethe-Bloch curves for different particles species. Right: Momentum resolution in the TPC as function of transverse momentum $p_t$. The insert shows an example of a cosmic track used for the calibration. }
\label{dEdx_momentum}
\end{figure}

\begin{figure}
\centering
\includegraphics[width=12cm,clip]{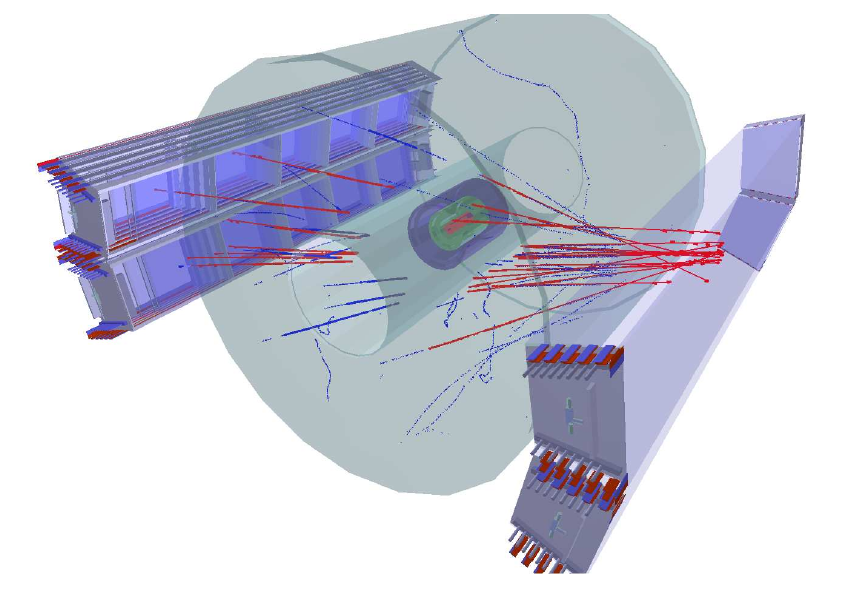}
\caption{A cosmic-ray event in the TPC taken with TRD L1 trigger.The four TRD super-modules are also shown.}
\label{TRDCosmic}
\end{figure}

\begin{figure}
\centering
\includegraphics[width=14cm,clip]{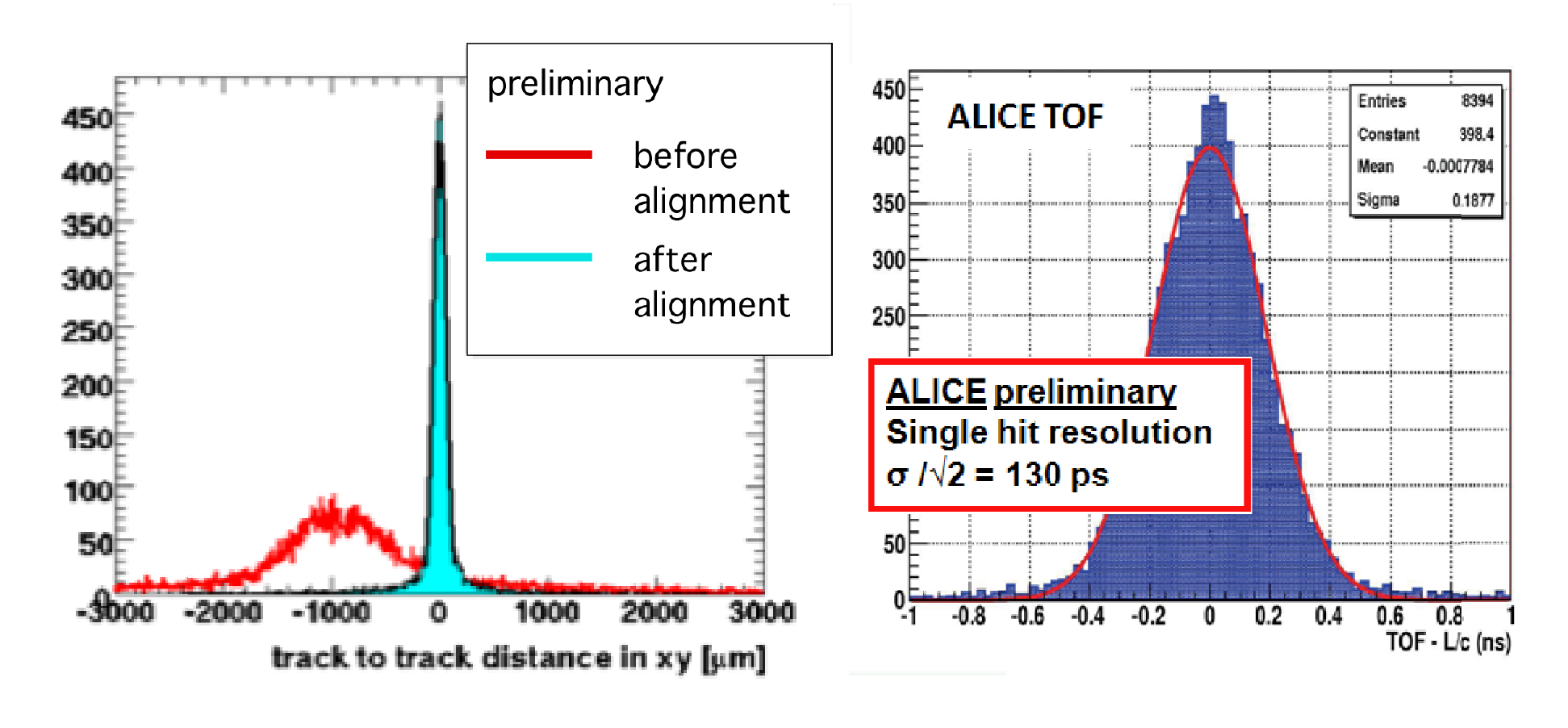}
\caption{Left: Preliminary resolution of the SPD after alignment with cosmic tracks. We compare the upper and lower halfs of cosmic tracks reconstructed in the SPD. Right: Histogram of the measured flight time between two TOF modules. We subtract L/c, with L being the length of the trajectory, and c is the speed of light}
\label{ITS_TOF}
\end{figure}

\section{Detector Commissioning}

First particles from the LHC were seen on June 15$^{th}$, when the ALICE Silicon Pixel Detector received a significant load of muons after the injection of particles from the SPS into the LHC. During LHC tune-up until October 2008 the detectors were mostly operated under "realistic" conditions, i.e., continuous 24 hours data taking in global mode (i.e. all detectors participating in the data stream). The data taken were either comics, laser,  source or pulser data and are used to calibrated and align the detectors.

\subsection{Time Projection Chamber Calibration and Commissioning}
\paragraph{\bf Gain Calibration and dE/dx calibration}\noindent
The gain of 557,568 individual readout pads located at the two TPC endplanes was calibrated injecting radioative $^{83}Rb$ into the TPC gas. $^{83}Rb$ ($\tau_{1/2}$=86.2 d) decays to $^{83m}Kr$ via electron capture. The decay of the isomeric Krypton state ($\tau_{1/2}$=1.83 hrs) produces via IC with a large probability an electron cluster of 41 keV, which is used to equalize the gains. This calibration procedure reveals an average pad-to-pad gain varation of $\pm15\%$, which is well within the range expected from the production tolerances. 
Having the gain equalized,  the response of the pads to the differential energy loss dE/dx of charged tracks in the gas is calibrated as a function of  $\beta\gamma$ to match the Bethe-Bloch curves (cf. Figure ~\ref{dEdx_momentum}). The present resolution in dE/dx is of the order of 6\%, which is close to the design value of 5.5\%. This allows to separate particles (protons, kaons, pions and electrons) on a statistical basis in the region of the relativistic rise, i.e., at momenta above 1-2 GeV/c.

\paragraph{\bf Temperature Stabilization, Space Point and Momentum Resolution}\noindent
The strong dependence of the drift velocity on the temperature for the gas mixture used in the TPC ($1/v_{drift} \times dv_{drift}/dT = 0.35\%/K$) requires to stablize the temperature to better than $0.1 ^\circ C$ over the full drift volume of $88 m^3$. This is achived by very efficient heat removal from the front-end electronics and by well defined iso-thermal surfaces around the TPC barrel via thermal screens. In addition, the Al-bodies of the readout chambers and the resistive voltage divider chain of the field cage ("resistor rod"), which face or are inside of the gaseous volume, respectively, are set to precisely defined temperatures. The cooling system consists of 59 individual sub-atmospheric cooling circuits, which can be regulated in flow and temperature. 
Both the space point and the momentum resolution of individual tracks are determined from cosmics rays showers measured in the TPC. The resolution relevant for high $p_t$ tracks (large drift) is determined from cosmic rays and is of the order $300-800 \mu m$. These values are in accordance with the design specifications.
The momentum resolution has been estimated from the correlation of semi-tracks, i.e., cosmic tracks crossing the TPC through the center have been split into to half tracks. The extracted momentum resolution is shown in Figure~\ref{dEdx_momentum} and is $\approx 6.5\%$  for 10 GeV/c tracks  and 1 \% for 1 GeV/c tracks. This is, with the present set of corrections, somewhat worse than the design resolution (4.5\% at 10 GeV/c).

\subsection{Transition Radiation Detector Commissioning}
Four supermodules were present in the ALICE setup in 2008 with the full data readout chain and participated in the cosmic-ray data taking. A display of a cosmic-ray event is shown in Figure~\ref{TRDCosmic}. 50 000 horizontal tracks were acquired  employing the TRD Level 1 trigger. In a first iteration the calibration parameters  for gain and drift velocity were determined and the cluster resolution for tracks crossing several chambers was determined. When the chamber misalignment is accounted and corrected for, the resolution is of the order of $300 \mu m$, i.e., within the design specifications.

\subsection{Inner Tracking System}
The Inner Tracking System consists of three different silicon detector technologies with two layers each: high resolution silicon pixel (SPD), drift (SDD) and strip (SSD) detectors. The system is fully installed and commissioned. The detectors were calibrated using cosmic ray samples of $10^5$ events. Both the Millepede method (default method, as for all LHC experiments) as well as an iterative approach were used to determine the aligment parameters. In addition, a dE/dx calibration in the SPD and SSD was performed. As an example, we show in Figure~\ref{ITS_TOF} the preliminary result of the alignment procedure for the SPD. The residual misalignment in the SPD is better than $10 \mu m$ yielding an impact resolution $\sigma = 56 \mu m$. These values are close to the design values.

\subsection{Time-of-Flight}
The nominal overall time-of-flight resolution for the TOF detector is $\approx 120~ps$. However, this requires a precise $t_0$ determination, i.e.  slewing corrections, knowledge of PCB  trace and cable length variations , etc. These cannot be obtained from cosmic data, but  require at least 1 week of  p-p data at  $L= 5 \times 10^{29} cm^{-2} s^{-1}$. The present resolution without final $t_0$ calibration,  is, however, already $\sigma _{MRPC} = 185 ps /\sqrt {2} \approx  130 ps$ (cf. Figure~\ref{ITS_TOF}) , which promises a value after a final calibration close or better than the design resolution.

\section{Summary}
Where possible the detector performance was evaluated using cosmic muons and the first few beam particles delivered by the LHC. All installed detectors are fully commissioned and shown to be performing close to their specifications. The current shutdown period 2008/2009 is used to install additional detectors. The ALICE experiment is ready for recording the first proton-proton collisions in the LHC, expected in late 2009, and the collaboration eagerly awaits the first heavy-ion collisions at the end of the upcoming run.

\vspace{1 cm}

\bibliography{LLWIproceedings_Schmidt}
\end{document}